\documentclass[pra]{revtex4-2}
\usepackage{graphicx}
\usepackage{xcolor}

\begin{document}

\title{Ionization of Rydberg atoms embedded in Ultracold Plasma due to electron-atom interaction}
\author{Satyam Prakash, Ashok S Vudayagiri}
\affiliation{ School of Physics, University of Hyderabad, Prof.C.R. Rao road, Gachibowli, Hyderabad 500046, \bf{India}}

\begin{abstract}
When ultracold plasma is generated using photonization of laser cooled atoms, some atoms reach only upto Rydberg states. These in turn interact with the free electrons of the plasma and get ionized further. We study
the interaction of electron-Rydberg atom using potential scattering technique in quantum mechanical domain and compute the associated cross sections for Cesium atoms, analytically. We notice a close agreement with the experimental data of ionization of Rydberg atoms as reported in Phys. Rev. A 71, 013416 (2005). The experiments showed a rapid increase in ionization above a specific Rydberg state. Our theory supports the same, and also indicates that this is due to the relation between scattering length and the radius of the orbit. 
\end{abstract}

\maketitle


\section{Introduction}
 Ultracold plasma(UCP) is experimentally obtained by resonant photoionization of laser-cooled atoms. Since the resonant processes do not provide additional kinetic energy, the plasma thus created is still at a very low temperature. The ions are typically around the temperature $\sim$10 $\mu$K, whereas the electrons have a substantially higher temperature of the order of $\sim$100 mK \cite{1}. On the other end of the temperature spectrum are the high-temperature plasma which are also known as High Energy Density Plasmas (HEDP). Their behavior has been extensively studied as they play a key role in military applications and nuclear fission research and technology. Though UCPs and HEDPs are radically different in density and temperature, their characteristics overlap for low screening strength($\kappa$) and low coupling constant($\Gamma$)\cite{11,2,3} which paves the way for multi-scale theories at the interface of atomic physics, plasma physics, and condensed matter physics. As a result, well-diagnosed experiments on ultracold plasma and its theoretical understanding can provide insights into ionosphere dynamics, and internal mechanisms within the core of white dwarfs and gas-giant planets. 
 
 Several research groups have modeled the evolution of ultracold plasma using hydrodynamical equations which could explain phenomena such as ultrafast electron cooling, plasma expansion, and the appearance of strongly correlated regime \cite{11,3}. Most of these  consider the UCP as a dilute plasma since the mean inter-atomic distance is large enough and the interactions between the constituents are very few.  On the other hand, hot plasmas have been analysed using  the classical, non-equilibrium thermodynamical approach and the basic plasma properties such as  quasineutrality, Debye shielding, and plasma oscillations are all explained from thereon.  Similarly, extending this classical approach to UCP has provided insights into some of their collective behavior such as electron-electron correlation, the temperature dynamics of the electrons, and the properties of expanding plasma cloud etc. \cite{3,4,5}. A few other properties require a fully quantum treatment. 

 One feature commonly encountered in UCP is the presence of neutral atoms which are not ionized but reach high-lying Rydberg states instead. They are produced mainly due to the tail end of the spectrum of the laser pulse where the energies are not sufficient enough to cause ionization. When conditions are right, these Rydberg atoms interact with other Rydberg atoms and also with electrons from ionized atoms. These interactions causes the Rydberg atoms to ionize \cite{4,5}, resulting   an increase in plasma density over and above the initial density.  The phenomena are explained either using a fully classical approach using Coulomb interaction \cite{5} or a semiclassical approach which accounts for quantum effects such as screening and diffraction. It is important to note that in the case of both moderately hot plasmas as well as UCPs, quantum effects become important due to the relatively long interaction times($\sim$100 $\mu$s) between ions and electrons, which needs to be accounted for. With this purpose, we investigate the quantum nature of the interaction of free electrons with Rydberg atoms and compute the probability of ionization of the Rydberg atoms.

 In the semi-classical approach, a classical potential function including the screening and diffraction effects of a multicomponent plasma is utilized to solve the Schrodinger equation \cite{6,7,8}. The screening part was suggested by Redmer and Ropke \cite{8} by calculating the effect of surrounding ions in reducing the Coulomb interaction. They added screening terms to the Hartree-Fock potential, polarisation potential as well as exchange potential. The total sum of these potentials is called optical potential which was used later by several other groups to calculate the total electron-atom scattering cross-section for alkali plasmas and noble gas plasmas\cite{6,7}.  The optical potential approach has been so far used only for moderately hot plasmas but not for UCPs. We will show that the quantum approach, using this same optical potential,  will give better agreement with experimentally observed values. Compared to Moderately hot plasma, UCP has a longer Debye radius and hence a smaller screening parameter($\kappa$) along with a smaller degeneracy parameter. The effect of these parameters is crucial for the study of electron-atom interaction in the UCP.

  Experiments by Vanhaecke et. al have shown that a dense sample of Rydberg atoms spontaneously evolves into a plasma due to Rydberg - electron interactions \cite{4}. A theoretical explanation for this was provided by Thomas Pohl using the classical, Monte Carlo simulation of evolution of electron temperature\cite{5}. We use quantum mechanical scattering and show that it plays a significant role, especially at lower temperatures($\sim 1mK$) and larger interaction times($\sim 100\mu s$). We analytically calculate the appropriate optical potential for the Rydberg atoms, mainly using the screened polarisation potential and compute the cross sections for their ionization due to electron scattering.  The polarisability of the Rydberg atoms has been taken care of using the static non-relativistic dipole polarisability for hydrogen-like species in the (n,l) state \cite{9}. Our results agree qualitatively and quantitatively with the experimentally measured values of the percentage ionization of Rydberg atoms as measured by Vanhaecke et al. These experiments also indicate that the probability of ionization are much lower when the atoms occupy energy levels lower than n=30 compared to when the value of n is much higher. Our calculations also show the reason for this behavior.

This communication is organized as follows (i) We first provide an overview of existing theoretical models on electron-atom scattering within a plasma medium (ii)identify the conditions of UCP under which the potentials used by us are valid (iii) calculate the full form of potentials for Rydberg atoms embedded within a UCP (iv) obtaining the cross section for electron scattering from a single Cesium atom, which causes it to ionize and (v) finally multiply with electron and atom densities to compare with the experimentally measured values of `percentage ionization'.

\section{Electron-atom interaction potential in UCP medium}
At first, we compute the potential for electron-atom interaction, extending the work of Ramazanov et al, who calculated the electron-atom interaction in a partially ionized plasma medium using the dielectric function method \cite{7}. Redmer and Ropke introduced an exponential factor into this to account for the long Coulomb interaction to 

An exponential factor was introduced into this by Redmer and Ropke to account for the long-range Coulomb interactions and overcome its divergence at shorter and longer distances. This modification to the Hartree-Fock potential is based on the interaction of electrons with the unperturbed field of the atoms. To calculate the Hartree-Fock potential, we consider the Hamiltonian for the electron wavefunction, given by:
 $$
\left(-\frac{\nabla^2}{2}- \sum _ {i=1}^ {m}\frac{Z_i}{|\vec R_i-\vec r|}+ \sum _ {i=1}^ {n}  \int  {d\vec r_i} \frac{|\psi_i(\vec r_i)|^2}{|\vec r _i-\vec r|} \right)\psi_j(\vec r)=E'\psi_j(\vec r)$$
The first and second terms in the above equation are the kinetic and the potential energies due to nuclear attraction. The third part is the Hartree term which is the potential seen by the electron moving in the potential generated by the remaining electrons at {$\vec r_i$}.

The total electronic wavefunction can then be written as the product of individual electronic wavefunctions:
  $$\psi(\vec{r_1},\vec{r_2},\vec{r_3},...\vec{r_n})= \prod_{i=1}^{n} \psi_i(\vec{r_i}).$$ 
Here, $\rho(r)$ is the electronic density which is integrated over the whole volume of the atom. To solve the equation above, the Hartree-Fock potential between the incoming electron and the atom is given by the Coulombic interaction of the projectile electrons with the atomic nucleus and the shell electrons:

$$
V_{HF}(r) = \frac{e^2}{4\pi \epsilon_o}\left[-\frac{Z}{r}+\int{\frac{1}{|{\vec r- \vec r_1|}}}\rho( r_1)d^3r_1 \right] $$
For the plasma medium, the Coulomb interaction is replaced by a Debye potential following Karakhtanov et. al., as computed for a partially ionized Hydrogen plasma. By expanding the electron-electron repulsion part into radial and angular coordinates, Rosmej et al. derived the screened Hartree-Fock potential in terms of three integrals \cite{8}.

\begin{eqnarray*}
V_{\mathrm{HF}}^{\mathrm{s}}(r) & = & \frac{e^2}{4 \pi \epsilon_0}\left[-\frac{Z e^{-\kappa r}}{r}+I_1+I_2+I_3\right]	\cr
&&\cr
&&\cr
I_1 & = &\frac{e^{-\kappa r}}{\kappa r} \int_0^r 2 \pi r_1 \rho\left(r_1\right) e^{\kappa r_1} d r_1 \cr
&&\cr
&&\cr
I_2 & = &-\frac{e^{-\kappa r}}{\kappa r} \int_0^{\infty} 2 \pi r_1 \rho\left(r_1\right) e^{-\kappa r_1} d r_1 \cr
&&\cr
&&\cr
I_3 & = &\frac{e^{\kappa r}}{\kappa r} \int_r^{\infty} 2 \pi r_1 \rho\left(r_1\right) e^{-\kappa r_1} d r_1	
\end{eqnarray*}

The potential calculated above depends on the atomic number Z, the screening constant $\kappa$ for the given plasma medium, and the electron density distribution $\rho(r)$ specific to the atom. For intermediate screening parameters we have $ 0 <\kappa a_0<1 $, where $a_0$ is the Bohr Radius, and $\kappa=\frac{1}{r_D}$ is the screening constant with $r_D$ being the Debye radius. The asymptotic form of the potential gives rise to a repulsive Hartree-fock potential.  At shorter distances, the effect of the plasma medium is ignored, and the purely Coulombic potential can be derived from Poisson's equation.
$$
V_{\mathrm{HF}}^{\mathrm{s}}(r)=\frac{Z e^2}{4 \pi \epsilon_0} \frac{e^{-\kappa r}}{r} \mathcal{C}_0\left[\left(\kappa a_0\right)^2+\mathcal{O}\left(\kappa^4\right)\right] $$$$
\mathcal{C}_0=\frac{Z^{-1}}{6} \int_0^{\infty}\left(\frac{r_1}{a_0}\right)^2 4 \pi r_1^2 \rho\left(r_1\right) d r_1
$$

The quasi-neutrality of the Ultracold plasma implies that the external field is negligible but the excitation of the atoms to Rydberg states gives rise to the polarisation potential. In the plasma medium, the screened polarisation potential, as suggested by Redmer and Ropke by adding a screening factor to the Buckingham potential is 

$$V_{\mathrm{p}}^{\mathrm{s}}(r)=\frac{-e^2\alpha_p}{8 \pi \epsilon_0(r+{r_0})^4}{e^{-2\kappa r}}{(1+\kappa r)}^2 $$ 

For a quasi-neutral ultracold plasma that contains Rydberg atoms within, the static non-relativistic dipole polarisability($\alpha_p$) for hydrogen-like atoms in the nl states is given by \cite{9}.
$$\alpha_p=\beta_{nl}^{z}= Z^{-4}[n^6+7/4n^4(l^2+l+2)][a_0^3]$$
Here, Z represents the atomic number, and n, and l are the principal and azimuthal quantum numbers respectively.

The local exchange potential is calculated following the approach of Mittleman and Watson under free-electron-gas exchange approximation\cite {10}, where  $F(\eta)=\frac{1}{2}+\frac{1-\eta^2}{4\eta}\ln \left|\frac{\eta+1}{\eta-1} \right|$. Rosmej and Ropke \cite{8}  modified the exchange potential and obtained more accurate results within a phase shift error($<$ 0.1 rad) for slow-moving electrons.

$$V_{ex}^M(r,0)=-\frac{e^2}{4\pi\epsilon_0}\frac{2}{\pi}K_F(r) \hspace{1.5 cm} K_F(r)=[3\pi^2 \rho(r)]^{\frac{1}{3}} $$ 
$$K_{RRR}^2(r)= k^2+\frac{2m}{\hbar^2}[|V_{HF}(r)| + |V_p(r)|+|V_{ex}^M(r,0)]|$$

The theoretical methods mentioned above are used to compute the interaction potential for lower-lying states. However, in the case of atoms in the Rydberg levels both the electron densities and polarizabilities differ drastically, resulting in different behavior for polarization and interaction potentials. 

\subsection{A comparison between High energy density plasmas and Ultracold plasma} 

At first, we compare some of the characteristics of  High energy density plasmas (HEDPs) with those of Ultracold plasmas(UCPs). This is necessitated since we take a methodology initially developed to compute interaction potential for a moderately hot plasma and apply it to a much colder plasma. This approach requires identifying regimes of temperature and density where the quantum effects become relevant. The scaled screening strength $\kappa a_0$ and degeneracy parameter $\Theta$ for HEDP and low-temperature plasma, along with typical temperatures and electron densities are shown in Table 1 for comparison. The $\kappa a_0$ and $\Theta$ values are calculated from the electron temperature(T) and density($n_e$) values which have been theoretically extended from the known temperature and density values of HEDPs and UCPs. Degeneracy parameter is obtained from the relation $\Theta=k_BT/E_F$, where the Fermi Energy $E_F$ for a plasma is defined as $E_F=(\hbar^2/2m)(3\pi^2n)^{2/3}$. For high-density plasma, this reduces to the ratio between thermal de Broglie wavelength $\lambda_{dB}=\hbar/\sqrt{2m k_B T}$ to the Wigner - Seitz radius `$a$' as 

\begin{equation}
\Theta=\frac{k_BT}{E_F}\approx \left(\frac{\lambda_{dB}}{a}\right)^{-2}
\label{degeneracy}	
\end{equation}

\begin{table}[!h]
	\begin{center}
		\begin{tabular}{|cccc|}\hline
			\multicolumn{4}{|r|}{HEDP's - High temperature plasma \hspace{40pt} } \\ \hline 
			T(K) & 8 $\times 10^9$ & 8 $\times 10^6$ & 8 $\times 10^3$  \\ 
            $\Theta$ & 61.74  & 3.3  & 0.15 \\
	        n$_e$(cm$^{-3})$& 4.3 $\times 10^{26}$ & 1.1 $\times 10^{24}$ & 1.72 $\times 10^{21}$ \\		
			$\kappa$a$_0$ & 0.5 & 0.8 & 1 \\

			\hline
			\multicolumn{4}{|r|}{UCP's - Ultracold Plasma \hspace{90pt} } \\ \hline 
			T(K) & 0.8 & 0.08  & 0.008 \\ 
			$\Theta$ & 5.28  & 0.29  & 0.05 \\
			n$_e$(cm$^{-3})$& 1.72 $\times 10^{13}$ & 4.3 $\times 10^{13}$ & 1.72 $\times 10^{13}$ \\
			$\kappa$a$_0$ & 0.01 & 0.05 & 0.1 \\
			\hline
		\end{tabular}
		\caption{Comparison of the scaled screening strength($\kappa$a$_0$) and degeneracy parameters($\Theta$) between High Energy Density Plasma (HEDP) and Ultracold Plasma (UCP) for the given electron density$(n_e)$ and temperature(T). }
		\label{comparison_table}
	\end{center}
\end{table}

The HEDPs are typically at temperatures higher than $10^{3}$ K, with hottest ones having densities of the order $10^{26}$ cm$^{-3}$.  Plasmas, at temperature $\sim 10^{2}$ K are called moderately hot plasmas, they are partially ionized and therefore are not at equilibrium.  The high electron density in these plasmas, coupled with a high temperature allows for a very high energy density and subsequently very high pressure. Due to the high kinetic energy environment, the Coulomb coupling $\Gamma=Z^2e^2/4\pi\epsilon_o K_BT$ is usually low for HEDPs compared to the UCPs. Thus, HEDPs do not enter a strongly correlated regime as easily whereas UCPs on the other hand easily attain a high coupling regime ($\Gamma>1$). From Table 1, we can see that the high energy density plasma shows high screening strength and a very low value of degeneracy parameter [$\Theta=(\lambda_{dB}/a)^{-2}$] which indicates that the corresponding de-broglie wavelength of electron, $\lambda_{dB}=\hbar/\sqrt{2\pi mK_B T}$ is much larger compared to Wigner-Seitz radius $a=(3/4\pi n)^{1/3}$. For the degeneracy values($\Theta<1$), the HEDPs or more precisely the moderately hot plasma show quantum mechanical scattering. As can be noted from the table, HEDPs show this behavior for density($\sim 10^{21}$ $cm^{-3}$) and temperature ($\sim 10^4$ K). Increasing the electron density while keeping the temperature constant at 8$\times$ $10^3$K gives the degeneracy parameter less than 1, hence quantum mechanical scattering is suitable for these density and temperature values.

Screening strength $(\kappa a_o)$ is usually low for UCPs and the degeneracy parameter is assumed to be much higher compared to HEDPs. The UCPs are strongly coupled compared to the HEDPs. Although UCPs and HEDPs are radically different in density($n_e$) and temperature(T), their characteristics overlap for low screening strength($\kappa a_o$) and low coupling constant($\Gamma$)\cite{3}. Our calculations show that for a density of the order of $10^{13} cm^{-3}$, the UCPs have comparatively lesser degeneracy parameters for the electronic temperatures around 80mK and 8mK. It is worth mentioning that the theoretical parameters are closer to the actual UCPs in terms of temperature ($\sim$ 100 mK) but they differ from the actual UCPs in terms of density($10^{10}$ cm$^{-3}$). However, for the given values, a comparatively smaller degeneracy parameter of 0.05 means a de-Broglie wavelength that is approximately four times larger than the Wigner- Seitz radius and therefore will exhibit quantum effects in electron-atom scattering. If we can achieve similar densities($\sim 4 \times 10^{13} cm^{-3}$) at temperatures lower than 80mK, we get the UCPs with even lower degeneracy parameters for which the quantum mechanical electron-atom interaction will be valid.

\subsection{Potential for Rydberg atoms in UCP medium}
\begin{figure}[!h]
\begin{centering}
  \includegraphics[width=15 cm]{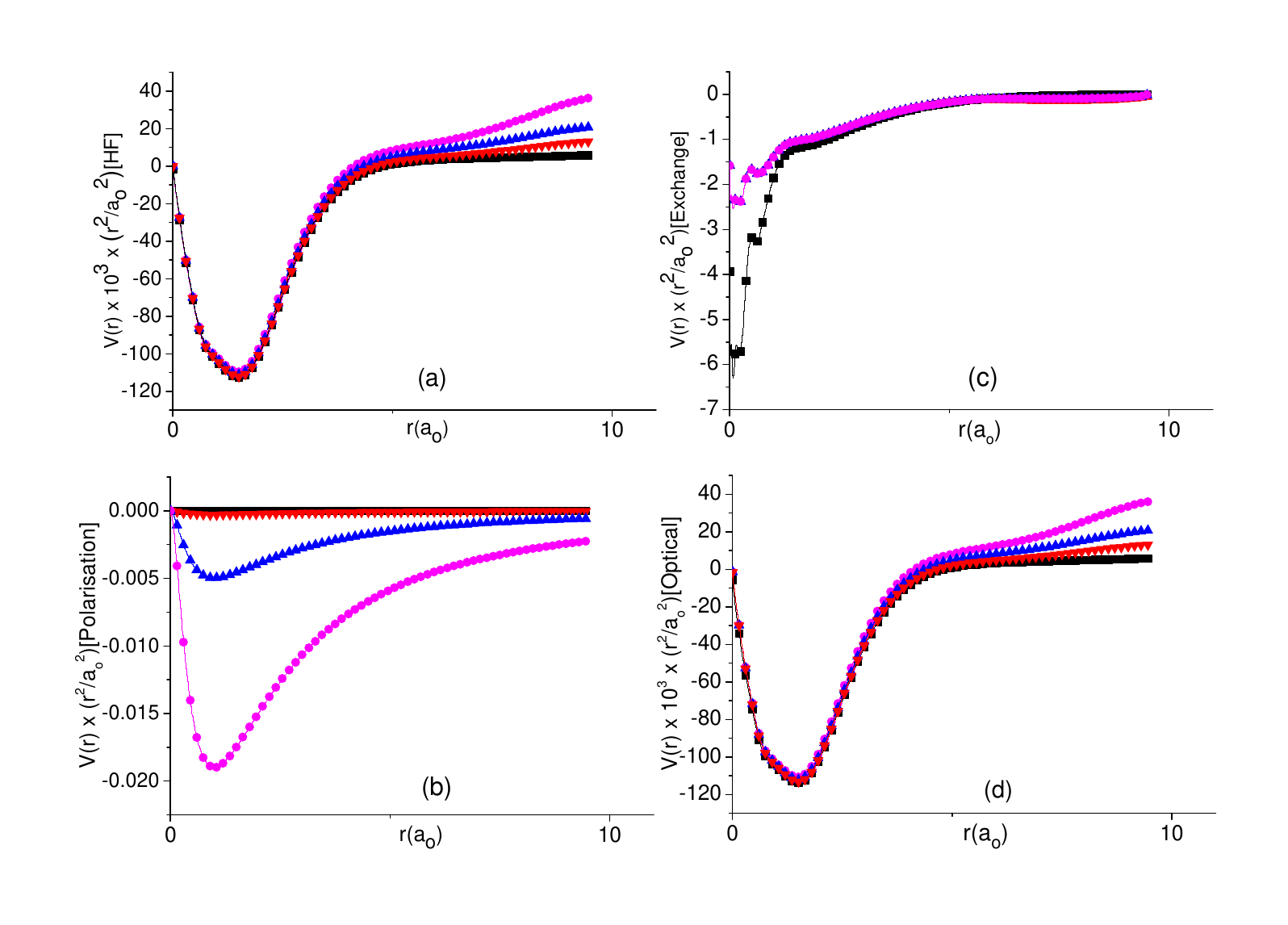}
  
  \caption{The interaction potentials (a)Screened Hartree-Fock potential[HF] (b)Screened polarisation potential (c)Exchange potential (d)Optical Potential calculated for Cesium ultracold plasma medium($\kappa a_o$ = 0.001). (color online) Black Squares are ground state, down triangle (Red) are n=20, up triangle (Blue) for n=32 and circle (pink) are n=40 }
  \label{potential}
\end{centering}
  
\end{figure}

In the subsection above, we have just discussed the regions where the interaction potential functions can be applied to UCP based on the values of the degeneracy parameter. Now, we compute the total optical potential using the electron densities for the Cesium atom for different Rydberg states based on the expressions for screened Hartree-Fock, Polarization, and Exchange potential obtained in the main section. The results for n=20, 32, and 40 are shown in figure \ref{potential}. Here the subfigures (a), (b), and (c) show the terms screened Hartree-Fock, screened polarization potential, and the screened exchange potential respectively. Figure (d) is the total of these terms which is called optical potential.

To obtain the Hartree - Fock component, at first the electron density function for Cs$^{+}$ in STO-3G basis was obtained from Atomic data tables \cite{12} for Cesium atom. The electron density for the excited state is computed using the Numerov method \cite{13}, for a given Rydberg state. These two are added to get the total electron density of the Rydberg atom. Hartree - Fock potential as well as the exchange potential is then calculated using the above electron density.  The polarisation potential depends on the shell number of the excited electron through the value of its non-relativistic polarisability given by the equation below.

\begin{equation}
\alpha_p=\beta_{nl}^{z}= Z^{-4}[n^6+7/4n^4(l^2+l+2)][a_0^3]
\label{polarization}
\end{equation}

It can be noticed that both the Hartree-Fock potential and exchange potential undergo small fractional change with a change in n. On the other hand, the polarisation potential increases to a large extent as the value of n increases. This is in agreement with the fact that the polarisation of the atom increases as $n^6$ when excited to shell level n \cite{9}.

In particular, the potential for Rydberg atoms - electron interaction is of our main concern for this work, which we obtain by computing the relevant Hartree-Fock, polarization, and exchange potentials and adding them to make the Optical Potential, as shown in the figure above.

\section{Cross-section calculation for Rydberg atoms placed in UCP medium}
Using the interaction potential of the Cesium Rydberg atom calculated above, we obtain the scattering cross section for electron - Rydberg atom scattering, which results in ionization of the Rydberg.

The partial wave analysis of electron-atom scattering describes the plane wavefunction of electrons as a sum of partial waves. The scattered wave function is given as the summation of scattering amplitude as a function of phase shifts $(\delta_l)$. These phase shifts are obtained as the asymptotic limit to the phase function $\delta_l(r)$ described by the phase equation \cite{14}.

\begin{figure}
  \begin{centering}
   \includegraphics[width=7.5 cm]{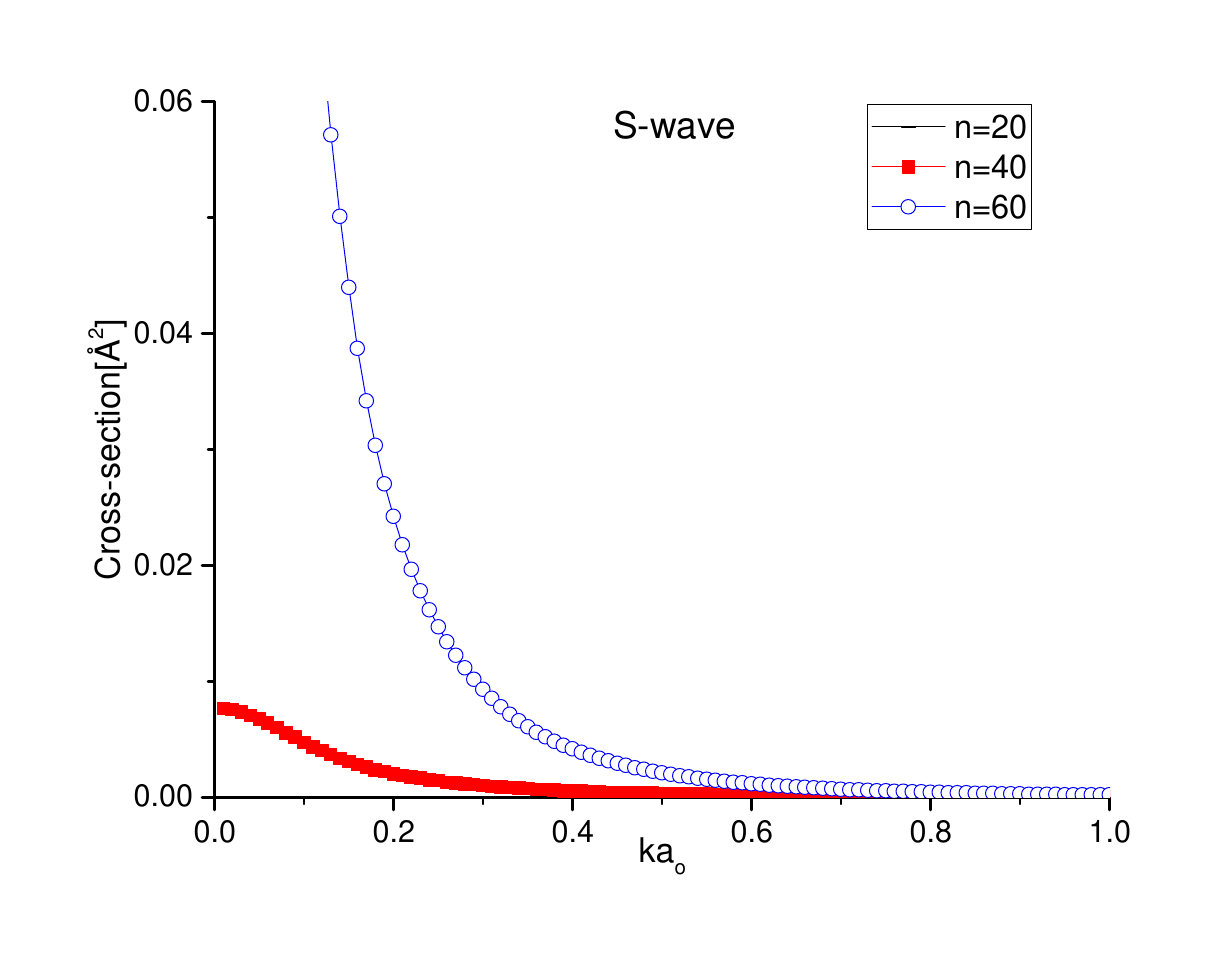} 
   \includegraphics[width=7.5 cm]{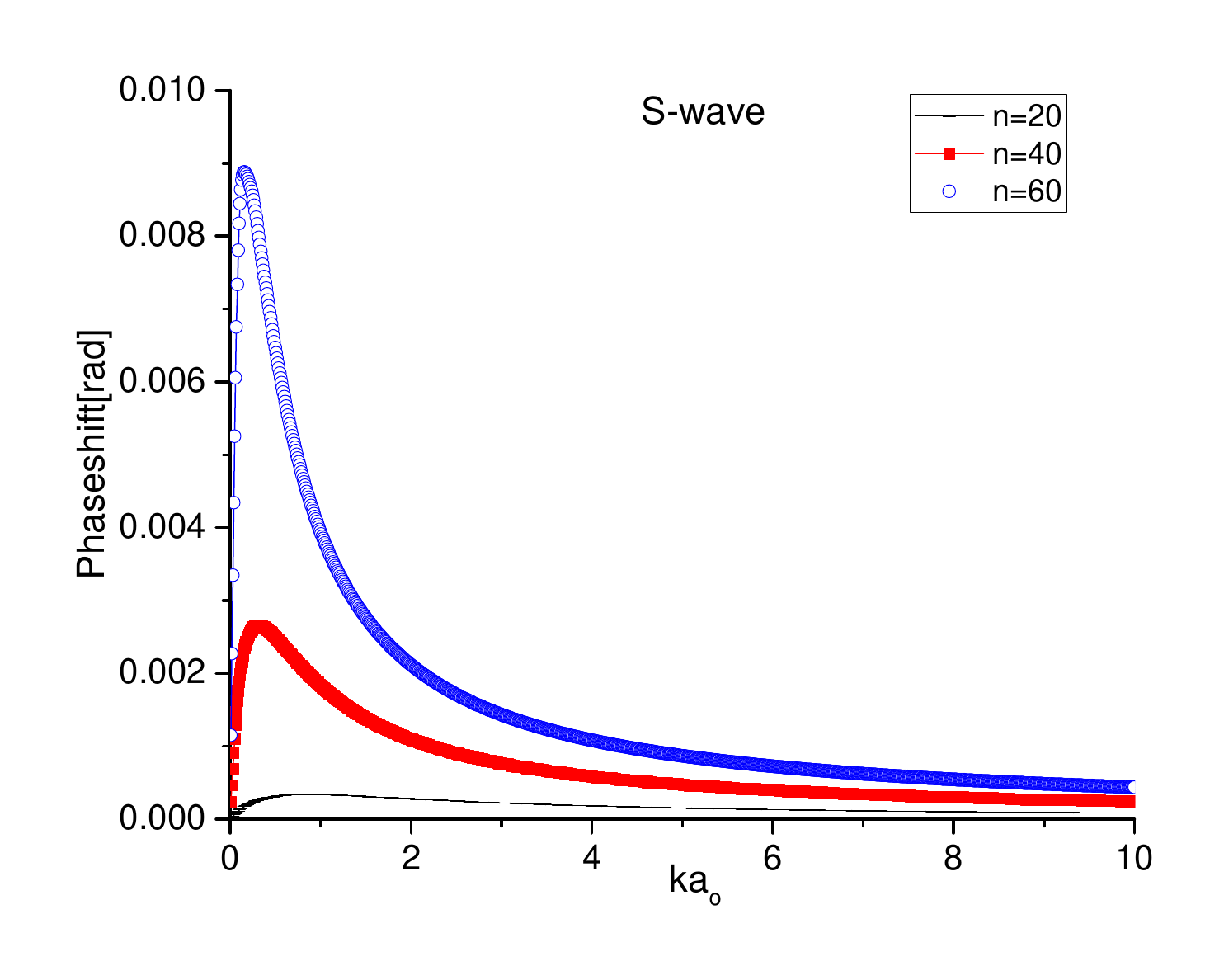}
  \includegraphics[width=7.5 cm]{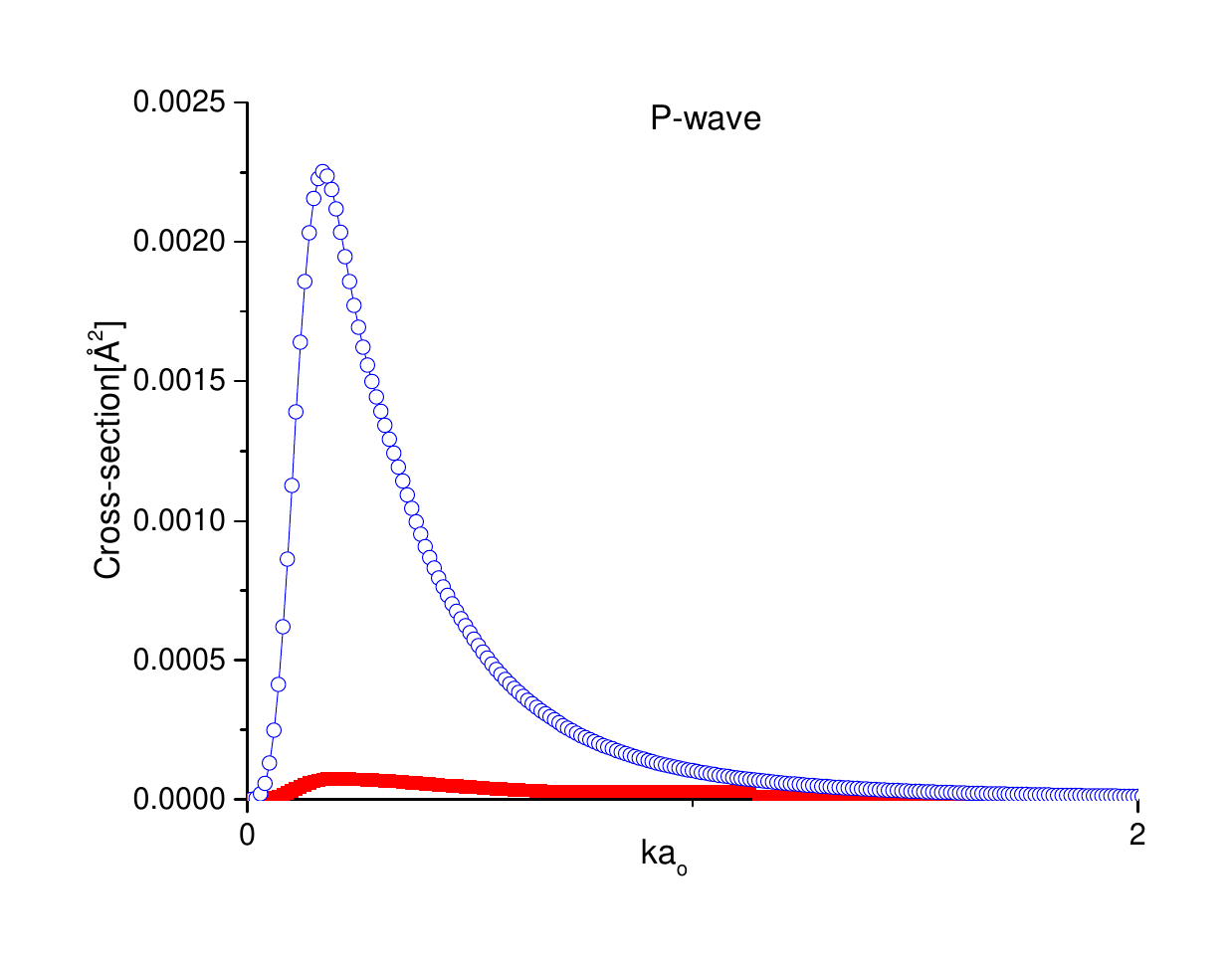}
   \includegraphics[width=7.5 cm]{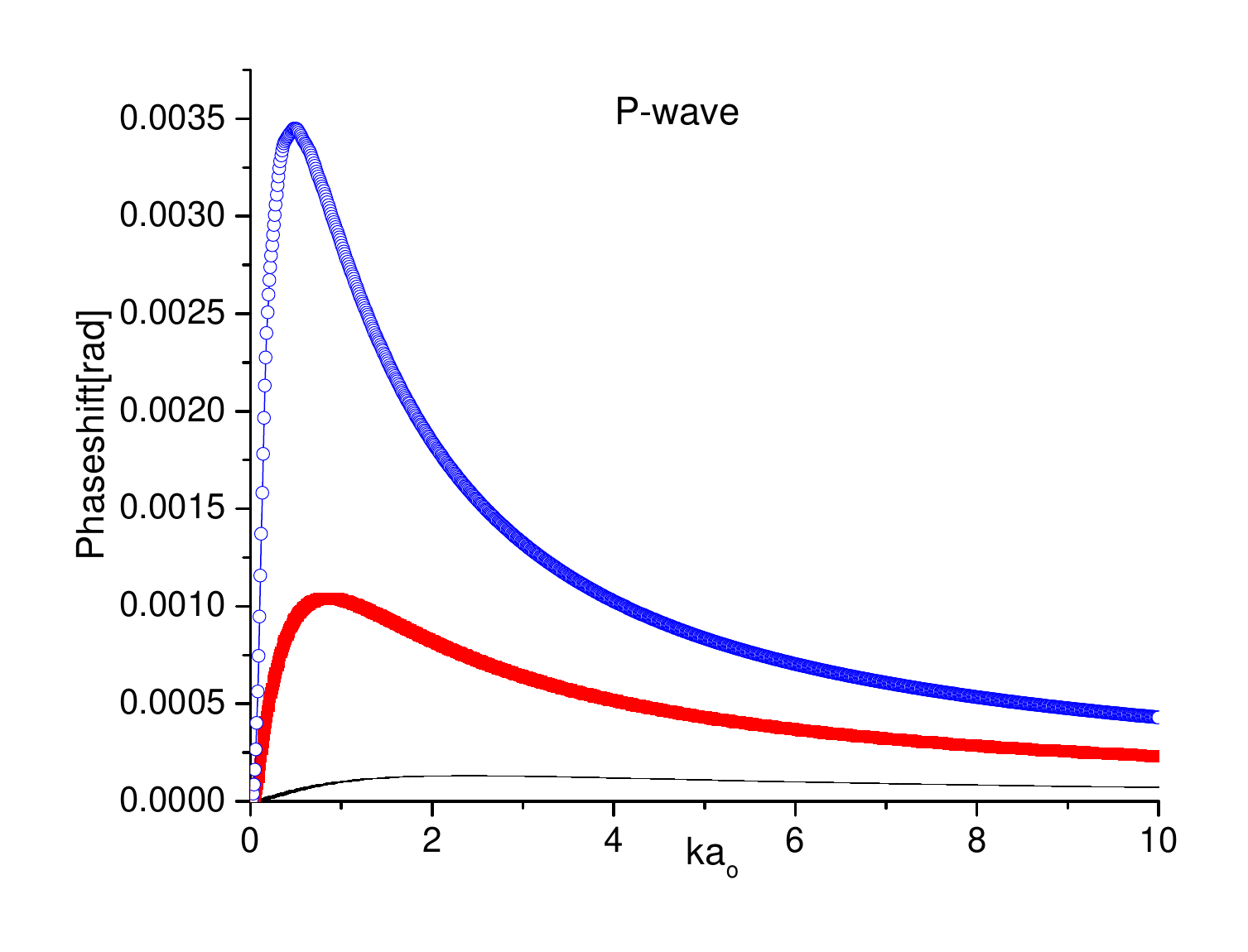} 
  \includegraphics[width=7.5 cm]{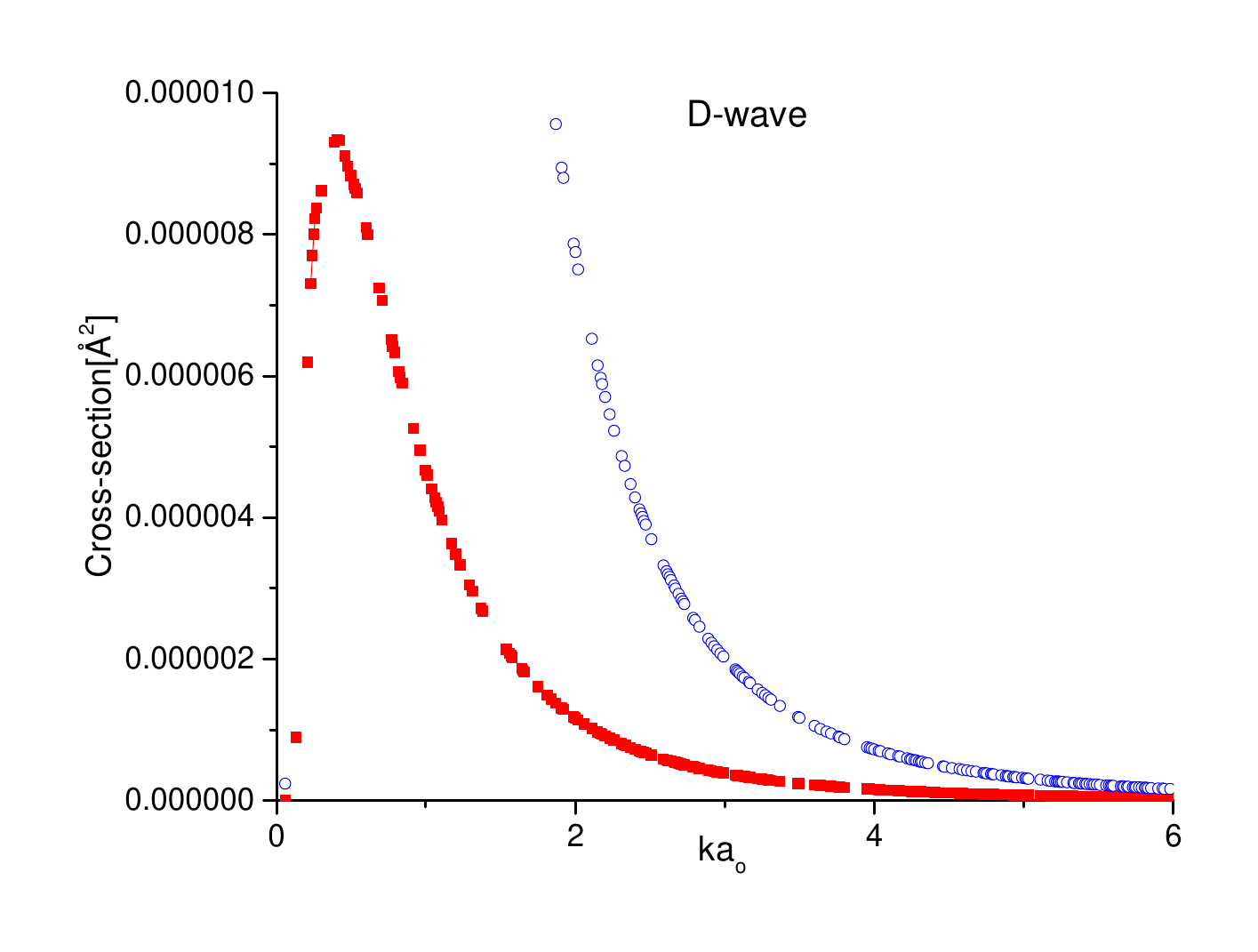}
  \includegraphics[width=7.5 cm]{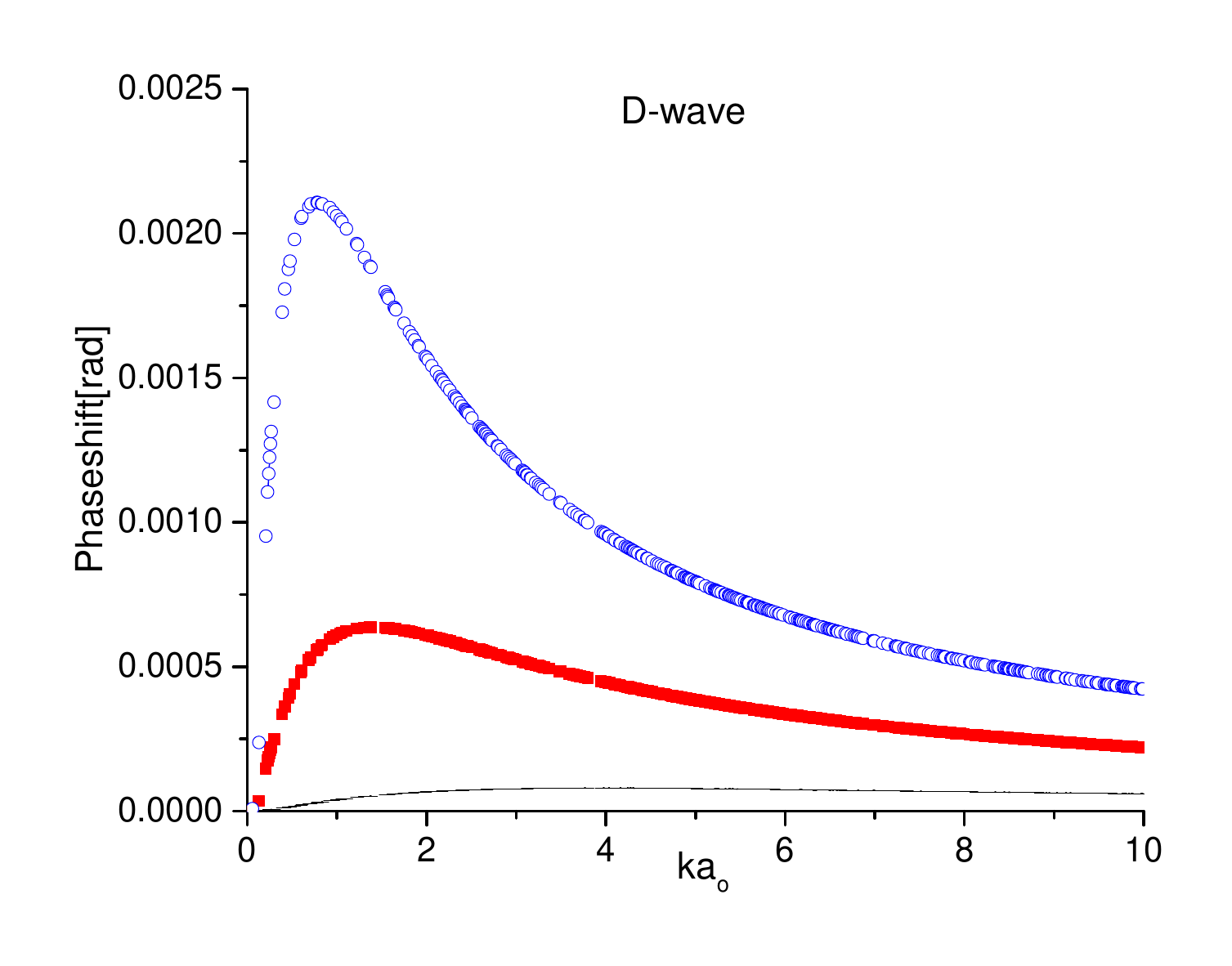} 
  \caption{Effect of Rydberg excitation on the phase shifts (left) and cross-section(right) plotted against electron momentum for electron-atom scattering in a plasma medium. Horizontal bars for n=20, closed squares for n=40 and open circles for n=60 }
  \label{phase_shift}
  \end{centering}
\end{figure}
\vspace{0.3 cm}

$$\psi_{in}= e^{ikz}=\sum_{l=0}^\infty(2l+1)i^lj_l(kr)P_l(\cos\theta)$$
$$\delta_l'(r)=-k^{-1}V(r)[\cos\delta_l(r)\hat{j}_l(kr)-\sin\delta_l(kr)\hat{n}_l(kr)]^2; \lim_{r \to \infty} \delta_l(r)=\delta_l$$ \\
The partial cross-section can be calculated using the phase method and the full cross-section is the sum of the cross-section over the distinct partial waves.
$$\sigma_l=\frac{4\pi}{k^2}(2l+1)\sin^2\delta_l \hspace{1 cm} \sigma_{total}=\sum_{l=0}^\infty \sigma_l$$

We have employed screened polarisation potential to study the Rydberg levels of the Cesium plasma as it simplifies the calculations.  The attractive polarisation potential for Rydberg atoms increases with the shell number. 

The effect of Rydberg levels on the phase shift and the cross-section has been plotted as a function of electron momentum (k$a_o$) in Figure \ref{phase_shift}. It can be noted that slow-moving(low-temperature) electrons in the plasma medium have a larger cross-section. Additionally, it is clear from the plots that s-wave cross-sections are dominant over the p-wave and the d-wave cross-sections. We observe that with increasing Rydberg levels, the phase shift due to polarization potential increases for all the partial waves i.e., the s-wave, the p-wave, and the d-wave. In other words, the cross-section increases for all the partial waves. However, the low phase shift for even higher Rydberg excitation shows the absence of bound states for excitation as high as n=100. The bound state formation as theoretically predicted by Levinson, was obtained for high-density Helium plasma \cite{6}. The existence of bound states in the UCP gives rise to unique atomic phenomena that need a separate discussion altogether. 

\section{Results and Discussion}
Spontaneous ionization of Rydberg atoms within a plasma \cite{15,16,17} is attributed to the interaction of electrons of plasma with Rydberg atoms. Vanhaecke et. al. \cite{4} studied the number of Rydberg atoms remaining unionized within the plasma, and showed its dependency on the power of the photoionizing laser. In other words, increasing laser power ionizes more atoms and hence an increase in electrons, which in turn increases the number of ionized Rydbergs, resulting in a decrease of remaining Rydberg atoms. A theoretical modeling for this process using Monte - Carlo technique was given by Pohl et.al. \cite{5}, and they explained the phenomena as a result of the avalanche process.

We have derived an analytical expression for the Rydberg atom-electron interaction potential, using a modified Hartree - Fock potential and computed the relevant cross-section for the case of Cesium atoms, present within a UCP.  The total cross-section would require integration over the Boltzmann distribution over all possible electron energies, computed as 

\begin{equation}
\sigma=\frac{4 \pi}{k^2}N \sin ^2 \delta_l \hspace{1 cm}  N= \frac{PN_o}{2P+c} 
\label{sigma}
\end{equation}

\vskip0.4cm

\begin{equation}
\sigma_{total} = \int{\sigma(v)n(v)dv}=\int{\frac{4\pi}{k^2}N\sin^2{\delta_l}~\frac{v^2}{\overline{v}^3}~e^{-(v^2/\overline{v}^2)}}dv
\label{sigma_total}
\end{equation}

\begin{figure}
  \begin{centering}
  \includegraphics[width=12.5 cm]{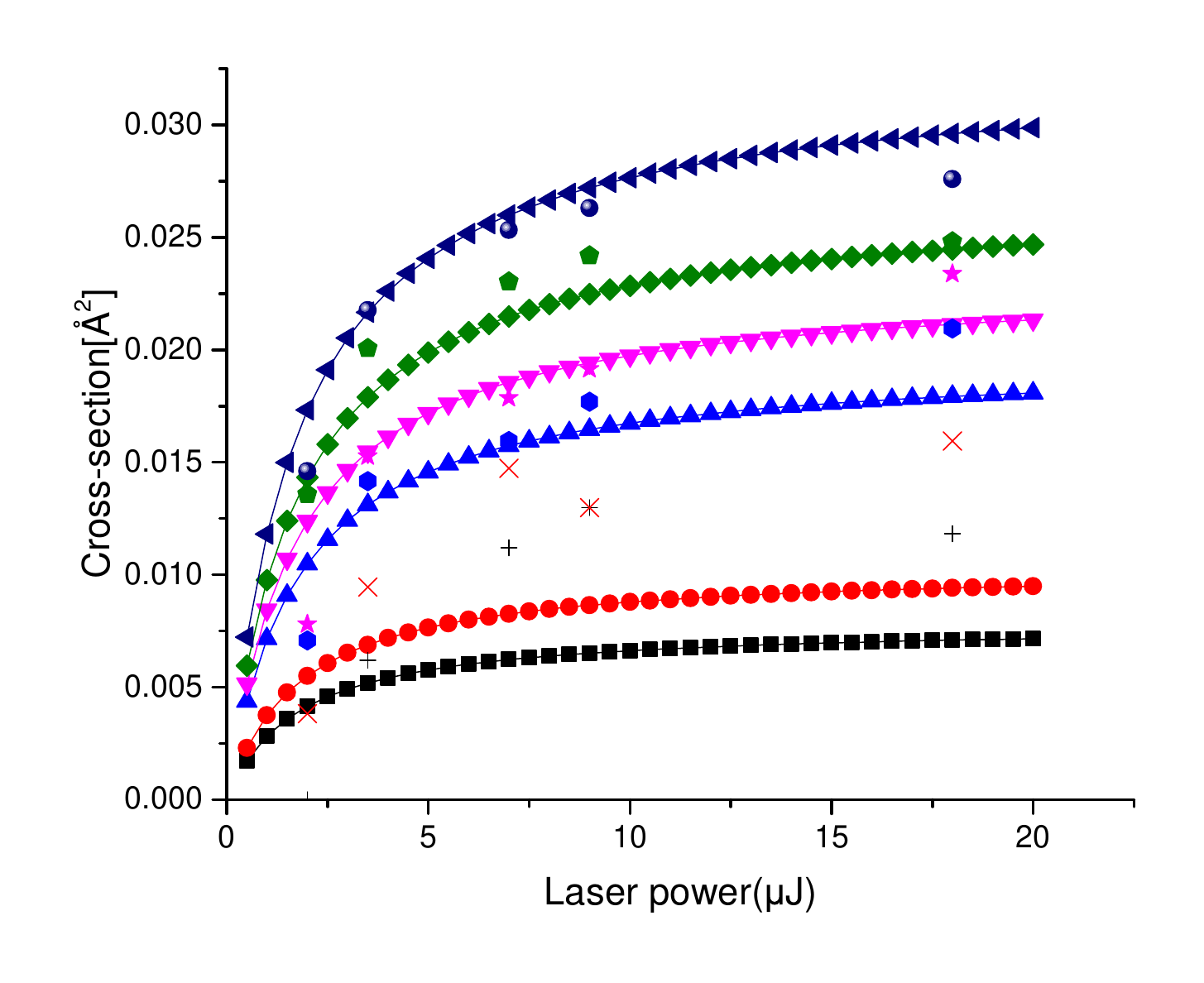}
  \includegraphics[width=7.5 cm]{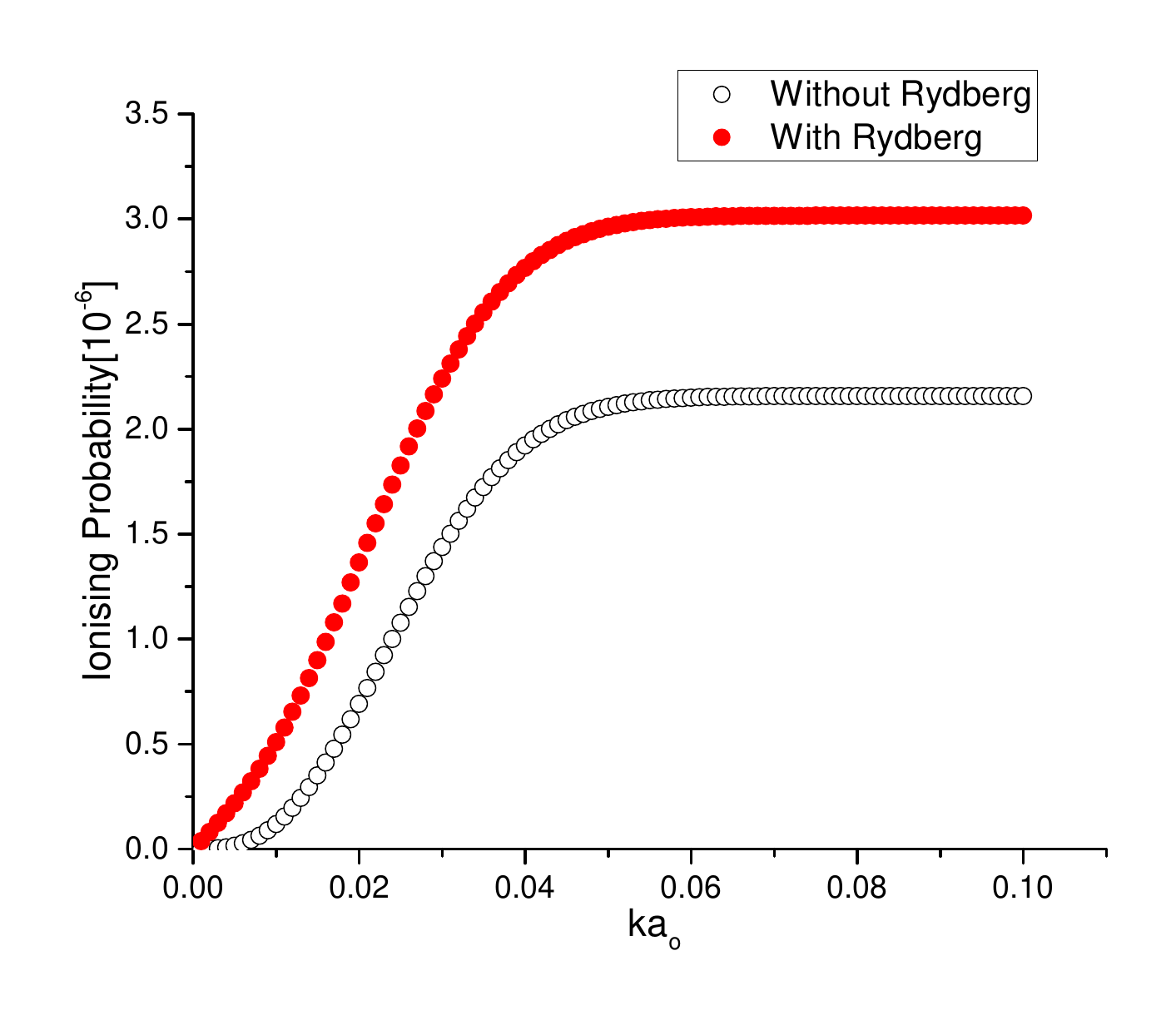}
    \includegraphics[width=7.5 cm]{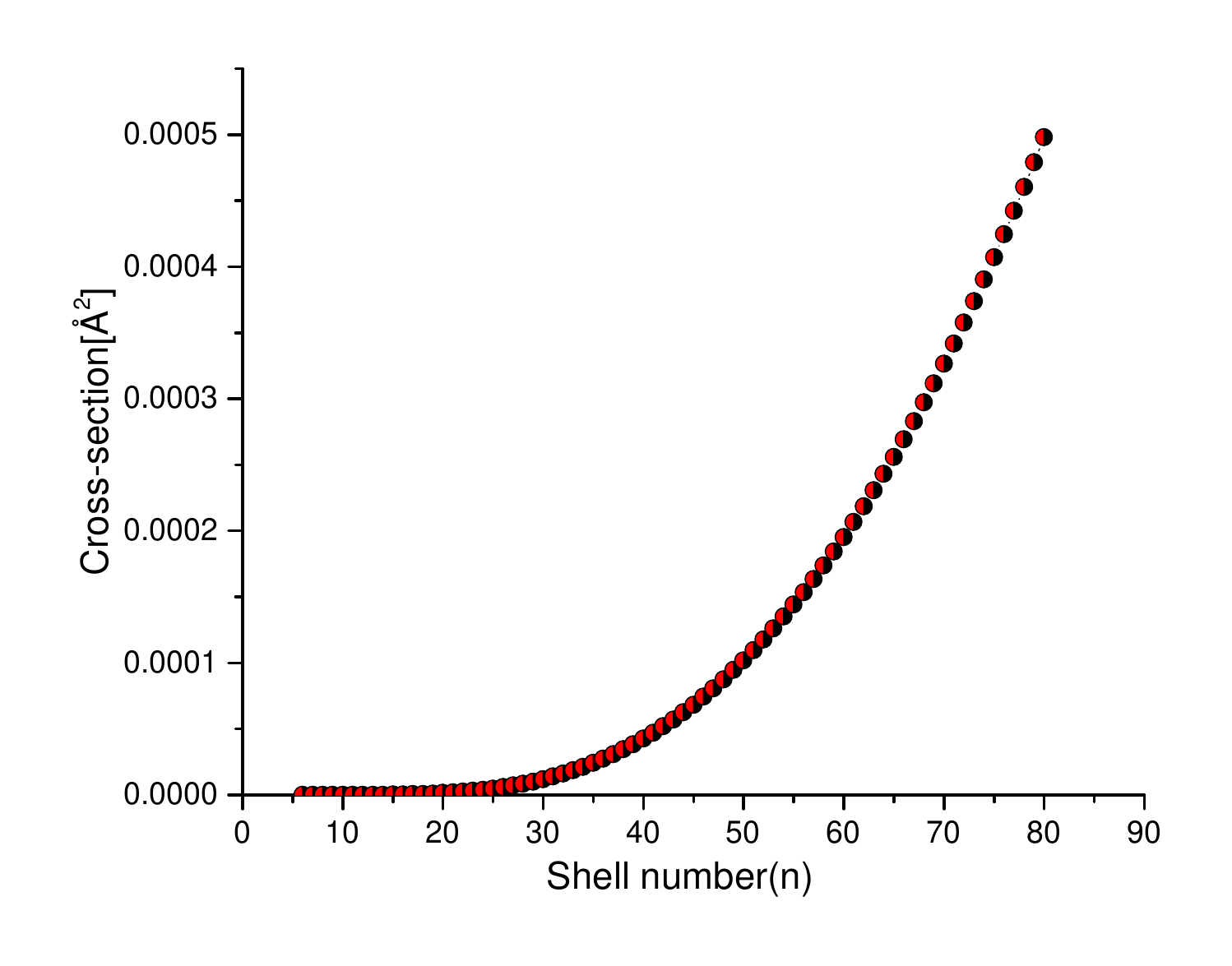}
  \caption{(a)Upper: Cross section for electron-atom scattering as a function of the power of ionizing laser for the shell numbers n=24(closed squares), 26(closed circles), 34(upward triangles), 39(leftward triangles); for shell numbers n=32(downward triangles), and 36(closed diamonds) as compared with experimental values of Vanhaecke et. al. and T. Pohl et. al. (b)Lower: Ionising probability as a function of electron momentum in the presence and absence of the Rydberg atoms(left). Variation of the cross-section with the excited shell number of the Rydberg atom(right).}
  \label{cross}
  \end{centering}
\end{figure}

Here $\sigma$ refers to the cross-section for a single electron-atom interaction $\sigma_{total}$ refers to the total cross-section for all electron-atom interactions, integrated over Maxwell - Boltzmann distribution for the given electron temperature.  k is the momentum parameter. $N_o$ is the initial number of electrons in plasma, taken to be of the order of  $10^6$ for our calculations. P is the laser power of the laser used for plasma formation, which determine the number of electrons. The quantity c in equation (\ref{sigma}) depends on the properties of the laser such as laser detuning, and saturation parameter,  and also on the specific atomic system\cite{18}. We use it as a constant in the fitting parameter for our data. The effect of temperature of the electron gas is incorporated as energies of electrons following the Maxwell-Boltzmann distribution, characteristic of the plasma \cite{19}. We normalize the electron numbers data from experiments into cross sections and compare them with the theoretically obtained values by us, as shown in the top portion of the figure (\ref{cross}). Plots with unconnected symbols are experimental data from reference \cite{4}, whereas plots with lines and symbols are our data. We can see that our calculated values are very close to the experimental value, although there is a small variation. Perhaps an experiment tuned to directly measure the interaction cross-section would be required to compare and find out if any additional physics needs that affect the cross-section need to be incorporated in the computation. 

 From Figure (\ref{cross}), we observe that the cross-section value increases with the laser power and also that a higher cross-section is obtained for higher Rydberg levels as the experimental values suggest. Experiments normally involve measuring electron current from the plasma. Hence, we computed the total electrons which can be measured either using a Faraday cup or a microchannel plate, by integrating over the Maxwell - Boltzmann distribution for a standard UCP. Multiplying the cross-section value by the number of Rydberg atoms will provide us with the number of additional electrons due to the ionization of Rydberg atoms.

\begin{equation}
P = \int{n(v)dv + \int{\sigma(v)n(v)dv}}  \hspace{2 cm}
n(v) = \frac{v^2}{\overline{v}^3}e^{-\frac{v^2}{\overline{v}^2}}
\end{equation}

The bottom plot(left) of Figure \ref{cross} shows a comparison between the number of electrons when Rydberg atoms are ionized (closed circles) and when they are not ionized (open circles). They indicate a significant increase in the number of electrons when Rydberg atoms are present. The bottom plot(right) shows a gradual increase in cross-section as the shell number of Rydberg atoms increases.

To compute percentage ionization as a function of shell number(n), the polarisability$(\alpha_p)$ has been chosen for the hydrogen-like systems and the cutoff parameter ($r_o$) using the analytical form suggested by Mittleman and Watson \cite{10}, which gives a form  
$$r_o=\left(\frac{\alpha_p a_o}{2z^{1/3}}\right)^{1/4}$$ 
where $\alpha_p$ is the polarisability, $a_o$ is the Bohr radius, and z is the atomic number of the Cesium atom. It was shown that this value of cutoff radius is relevant  to truncate the unphysical short-range contribution of polarisation potential \cite{20}. Based on this, our calculations show the cutoff radius is truncated up to the 5s(Inner shell) radius of the Rydberg atom only if the Rydberg is excited to n=30. At n=30, the cutoff radius approaches 1.8 Å which is also the inner shell radius of the Cesium atom. Hence, in Figure (3) we observe that the cross-section drastically increases beyond n=30. The experimental results affirm this calculation too \cite{4}. Furthermore, the cross-section radii that we have obtained using $r=\sqrt{\sigma}/\pi$ are of the order of the size of the Rydberg sample($\sim$ 0.1 mm) suggesting the high probability of quantum mechanical scattering-induced ionization of the Rydberg atoms.

Vanhaecke et. al. \cite{4} had attributed their experimental results to avalanche ionization due to collision between plasma electrons and Rydberg atoms. These observations agree well with our cross-section calculations too which indicate that much of the information is contained in the associated atomic processes as had been reported in multiple theoretical description of electron-atom interaction in the plasma environment \cite{5,21,22,23,24}. A characteristic of the avalanche process is that it depends upon the interaction time - a longer interaction time results in a higher percentage of electrons and ions.  This has been obtained in Figure 2, where we have shown that for low electron energies having a longer interaction time with the Rydberg atoms, the cross-section values are higher. Secondly, it is also observed that the percentage of ionization is proportional to the plasma density but is irresponsive to the Rydberg density ruling out the effect of autoionization of the Rydberg atoms. This agrees well with our cross-section calculations because the expression for total cross-section contains N, which is number of plasma electrons, which is directly proportional to the plasma density but not the density of Rydberg atoms. A contradicting observation was also made that up to 100 $cm^{-1}$ above the ionization limit, the percentage ionization did not depend strongly on the plasma electron energy which seems to contradict our model because in Figure 2 our simplistic assumption of the single electron-atom cross-section calculations depending on the energy of plasma electrons might not be the exact value for the total integrated cross-sections which determine the ionization efficiency for the complete system. 

\section{Conclusion}

We have shown that quantum mechanical scattering cross-section can be a new approach to studying the phenomena involving electron-Rydberg atom interactions. We show the consistency of our model with the experimental results and observe that the underlying atomic processes can be crucial to understanding the ionization of Rydberg atoms. More quantitatively, our calculations explain why ionization drastically reduces when the scattering length approaches the Bohr radius. For the further scope of this work, reference \cite{19} refers to an experimentally observed behavior of unusual loss of ions and deepening of the potential well that has been attributed to the possible formation of bound states. The inelastic scattering cross-section calculations for this phenomenon can shed light on interesting physics happening inside the ultracold plasma. 


\subsection{\textbf{Acknowledgments}}
The author acknowledges the generous support of CSIR-HRDG, Ministry of science and technology, government of India for the award of JRF/SRF (award number-09/0414(13706)/2022-EMR-I) throughout the course of the research.

\end{document}